\documentclass[aps,preprint,showpacs,amsmath,amssymb,superscriptaddress]{revtex4}
\usepackage{graphicx}

\begin{document}
\newcommand {\nc} {\newcommand}
\nc {\beq} {\begin{eqnarray}}
\nc {\eol} {\nonumber \\}
\nc {\eeq} {\end{eqnarray}}
\nc {\eeqn} [1] {\label{#1} \end{eqnarray}}
\nc {\eoln} [1] {\label{#1} \\}
\nc {\ve} [1] {\mbox{\boldmath $#1$}}
\nc {\vS} {\mbox{$\ve{S}$}}
\nc {\cA} {\mbox{$\cal{A}$}}
\nc {\dem} {\mbox{$\frac{1}{2}$}}
\nc {\arrow} [2] {\mbox{$\mathop{\rightarrow}\limits_{#1 \rightarrow #2}$}}

\title{$^{11}$Li $\beta$ decay into $^9$Li and deuteron within a three-body model}

\author{D. Baye}  
\email{dbaye@ulb.ac.be}
\affiliation{Physique Quantique, C.P. 165/82,  
Universit\'e Libre de Bruxelles, B 1050 Brussels, Belgium}
\affiliation{Physique Nucl\'eaire Th\'eorique et Physique Math\'ematique, C.P. 229, \\
Universit\'e Libre de Bruxelles, B 1050 Brussels, Belgium}
\author{E.M. Tursunov}
\email{tursune@inp.uz}
\affiliation{Institute of Nuclear Physics, Uzbekistan Academy of Sciences, 702132, Ulugbek, Tashkent, Uzbekistan}
\affiliation{Physique Nucl\'eaire Th\'eorique et Physique Math\'ematique, C.P. 229, \\
Universit\'e Libre de Bruxelles, B 1050 Brussels, Belgium}
\author{P. Descouvemont}
\email{pdesc@ulb.ac.be}
\affiliation{Physique Nucl\'eaire Th\'eorique et Physique Math\'ematique, C.P. 229, \\
Universit\'e Libre de Bruxelles, B 1050 Brussels, Belgium}

\date{\today}
\begin{abstract}
The $\beta$-decay process of the $^{11}$Li halo nucleus into $^9$Li and $d$ 
is studied in a three-body model. 
The $^{11}$Li nucleus is described as a $^9\textrm{Li}+n+n$ system in hyperspherical coordinates 
on a Lagrange mesh. 
Various $^9\textrm{Li}+d$ potentials are compared involving a forbidden state, 
a physical bound state, and a resonance near 0.25 MeV in the $s$ wave. 
With an added surface absorption, they are compatible with elastic scattering data. 
The transition probability per time unit is quite sensitive to the location of the resonance. 
For a fixed resonance location, it does not depend much on the potential choice 
at variance with the $^6$He delayed deuteron decay.  
The calculated transition probability per time unit is larger than the experimental value 
but the difference can be explained by a slightly higher resonance location and/or 
by absorption from the $^9\textrm{Li}+d$ final channel. 

\end{abstract}
\pacs{23.40.Hc, 21.45.+v, 27.20.+n, 21.60.Gx}
\maketitle
%
\section{Introduction}
The $\beta$ decay with emission of a deuteron, also known as $\beta$ delayed deuteron decay, 
was first observed for the $^6$He halo nucleus \cite{RBG90,BJJ93,ABB02}. 
The difficulty of the measurement led to conflicting results and raised a number 
of theoretical questions. 
The problems are related with the fact that the branching ratio is much smaller than expected 
from simple $R$-matrix \cite{RBG90} or two-body \cite{DL-92} model estimates. 

A semi-microscopic study of the process has been able to explain 
that the low value of the branching ratio is the result of a cancellation 
between contributions of the "internal" and "halo" parts of the Gamow-Teller matrix element \cite{BSD94}. 
A fully microscopic description of the $\beta$ decay of the $^6$He nucleus 
to the $\alpha+d$ continuum \cite{CB-94} then indicated that the cancellation could 
be more important than expected from the available data of that time. 
Both works emphasized that the $\beta$ delayed deuteron decay probes the halo 
up to rather large distances. 

These interpretations were confirmed in a recent study based on a three-body model \cite{TBD06}. 
It was shown that the cancellation requires that the $\alpha + d$ potential 
contains a forbidden state below the $^6$Li ground state 
in order to have the correct node structure of the scattering 
wave function. 
The cancellation is so strong in the Gamow-Teller matrix element for $^6$He that it requires 
an almost perfect balance between the internal and halo parts, which should be fortuitous. 
A similar effect is thus not expected for other halo nuclei possessing a $\beta$ delayed deuteron decay branch 
such as $^{11}$Li or $^{14}$Be. 

The most interesting halo nucleus is probably $^{11}$Li \cite{ZDF93}. 
Its two-neutron separation energy is particularly small: $300 \pm 19$ keV according to the atomic mass 
evaluation \cite{AWT03} or $376 \pm 5$ keV according to a recent preliminary result \cite{BAG05}.  
This nucleus can be considered as a $^9$Li core surrounded by two halo nucleons distant from each other 
by more than 6 fm \cite{MLO01} in agreement with theoretical expectations \cite{Ba-97}. 
It differs from $^6$He by the fact that its core does not correspond to a closed shell. 
The halo structure is understood as due to a virtual state in the $s$ wave of the $^9\textrm{Li}+n$ interaction 
\cite{TZ-94}. 
Another difference with $^6$He is that the core is unstable. 
Therefore many more $\beta$ decay channels are open. 
This complicates experiments but also offers many opportunities to test models \cite{MBG96,BFG97,BGG97}. 
Among the possible channels, the delayed deuteron $\beta$ decay remains especially interesting 
because this decay essentially occurs inside the halo and can probe its properties. 
In experiments however, the deuteron decays can not easily be separated 
from the delayed triton channel \cite{LDE84}. 
Anyway the $\beta$ delayed deuteron decay has been observed \cite{MBG96,BGG97} 
with a branching ratio of $(1.5 \pm 0.2) \times 10^{-4}$ \cite{BGG97}. 
This order of magnitude is consistent with predictions of a simple model \cite{OS-95} 
and of a model based on a limited hyperspherical-harmonics expansion \cite{ZDG95}. 
The analysis of a new experiment is in progress \cite{Ra-06}. 
It is thus timely to reexamine this process at the light of the recent 
knowledge gained on the $^6$He decay. 

In the present work, we describe $^{11}$Li as a $^9\textrm{Li}+n+n$ system with effective 
$^9\textrm{Li}+n$ and $n+n$ interactions treated in hyperspherical coordinates 
with the Lagrange-mesh method \cite{DDB03}. 
The final $^9\textrm{Li}+d$ state is modeled as a deuteron wave function 
multiplied by a scattering wave function deduced from a potential. 
Experiments have revealed the important role played by a resonance around 18 MeV 
in the excitation spectrum of $^{11}$Be \cite{BGG97}. 
Taking also that information into account, 
the transition probability per time and energy units will be calculated. 
The total transition probability is constrained with the branching ratio. 
They will be analyzed by comparison with the $^6$He decay with 
emphasis on the role of the node structure of the scattering wave functions. 

In Sec.~\ref{sec:model}, the $\beta$-decay model for the $^{11}$Li two-neutron halo nucleus 
into the $^9\textrm{Li}+d$ continuum is summarized. 
The potentials and the corresponding three-body hyperspherical and two-body scattering wave functions 
are described in Sec.~\ref{sec:pot}. 
The properties of the Gamow-Teller matrix element are studied in Sec.~\ref{sec:GTI}. 
In Sec.~\ref{sec:res}, we discuss the obtained numerical results and compare them with experimental information. 
Conclusions are given in Sec.~\ref{sec:conc}. 
\section{Model}
\label{sec:model}
The initial wave function is expressed as a bound-state wave function of 
the three-body $^9\textrm{Li}+n+n$ system with local $^9\textrm{Li}+n$ and $nn$ interactions. 
We neglect the spin of the core. 
The total orbital momentum $L$ of the three particles is assumed to be equal to the total 
spin $S$ of the neutrons as for $^6$He \cite{TBD06}. 
Jacobi coordinates, i.e.\ the relative coordinate $\ve{r}$ between the neutrons 
and the coordinate $\ve{R}$ of their center of mass with respect to the $^9$Li core, 
are necessary to calculate the overlap with the final scattering state.  
These coordinates are conveniently replaced by hyperspherical coordinates 
which involve five angular variables $\Omega_5$ and the hyperradius $\rho$. 
The wave function is expanded over hyperspherical harmonics depending on $\Omega_5$ 
and on the hypermomentum $K$. 
The coefficients in this expansion depend on the hyperradial coordinate $\rho$ and are expanded 
in Lagrange functions \cite{BH-86} (see Ref.~\cite{DDB03} for details). 

For the final scattering state, we assume an expression factorized into the deuteron ground-state wave function 
depending on $r$ and a $^9\textrm{Li}+d$ scattering wave function depending on $R$ derived from a potential model. 
We neglect the small $D$ component of the deuteron. 

The transition probability per time and energy units is given by \cite{BD-88}
\beq
\frac{dW}{dE}= \frac{m_ec^2}{\pi^4 v \hbar^2} G_{\beta}^2 f(Q-E) B_{\rm GT}(E),
\eeqn{1}
where $m_e$ is the electron mass, $v$ and $E$ are the relative velocity and energy in the center 
of mass system of $^9$Li and deuteron, and $G_{\beta}=2.996 \times 10^{-12}$ is the dimensionless 
$\beta$-decay constant \cite{Wi-82}. 
 The Fermi integral $f(Q-E)$ depends on the kinetic energy $Q-E$, available for the electron and antineutrino. 
The mass difference $Q$ between initial and final particles is given in MeV by 
\beq
Q = 3.007 -S_{2n}
\eeqn{1a}
as a function of the two-neutron separation energy of the halo nucleus. 
With the $^{11}$Li value $S_{2n} = 300 \pm 19$ keV from the atomic mass evaluation \cite{AWT03}, 
$Q$ is equal to $2.71 \pm 0.02$ MeV. 
However according to a recent remeasurement, the $^{11}$Li two-neutron separation energy 
becomes $S_{2n} = 376 \pm 5$ keV \cite{BAG05} leading to $Q = 2.63$ MeV. 
We shall first use the standard value and then consider the importance of this modification. 

Since the total orbital momentum and parity are conserved, only the $l=0$ partial scattering wave contributes. 
Hence, only the initial $L=S=0$ component of $^{11}$Li described with a spin 0 core 
can decay to $^9\textrm{Li}+d$. 
In order to allow the use of a complex optical potential for describing the scattering states, 
we generalize the formula of Refs.~\cite{BSD94,TBD06}. 
The final state is described by an ingoing scattering wave. 
At energy $E$, 
a partial wave $u_{E,l}^{(-)}$ of an ingoing scattering wave function is related 
to a partial wave $u_{E,l}^{(+)}$ of an outgoing scattering wave function by 
\beq
u_{E,l}^{(-)} (R) = (-1)^l u_{E,l}^{(+)*} (R).
\eeqn{2}
The outgoing radial scattering wave functions 
\beq
u_{E,l}^{(+)} (R) = e^{i\delta_l} u_{E,l} (R)
\eeqn{3}
are normalized asymptotically according to 
\beq
u_{E,l} (R) \arrow{R}{\infty} \cos\delta_l(E) F_l (kR) + \sin\delta_l(E) G_l(kR),
\eeqn{4}
where $k$ is the wave number of the relative motion, 
$F_l$ and $G_l$ are Coulomb functions \cite{AS-70},  
and $\delta_l(E)$ is the $l$-wave phase shift at energy $E$. 
The subscript $l = 0$ is understood in the following. 

The reduced transition probability can be written as  
\beq
B_{\rm GT}(E) = 6\lambda^2 \left| e^{i\delta_0} I_E(\infty) \right|^2,
\eeqn{5}
where $\lambda = -1.25$ \cite{DMD90}. 
The phase in front of $I_E$ does not play any role if the potential is real. 
The integral 
\beq
I_E (R) = \int_0^R u_E(R') u_{\rm eff}(R') dR'
\eeqn{6}
depends on a cutoff radius $R$ over the relative coordinate between the core 
and the center of mass of the nucleons. 
Only its value at infinity is physically relevant but it will help us 
to understand the physics of the decay process. 
This integral involves scattering wave functions $u_E(R)$ and depends thus on 
the $^9\textrm{Li}+d$ relative energy $E$. 
This integral also involves an effective wave function 
\beq
u_{\rm eff}(R) = R \sum_K \int_0^{\infty} Z_{K}(r,R) u_d(r) r dr, 
\eeqn{7}
where $u_d(r)$ is the deuteron radial wave function depending on the relative coordinate $r$ 
of the two nucleons. 
The sum runs over the values of the hypermomentum $K$ in the expansion 
of the initial bound state. 
The function $Z_K(r,R)$ is the radial part of the $K$ component 
with all angular momenta equal to zero 
in the expansion of the initial wave function. 
Its expression is given by Eqs.~(3) and (14) in Ref.~\cite{TBD06}) 
where however a normalization factor $[(A-2)/A]^{3/4}$ is missing. 
The results of Ref.~\cite{TBD06}) must be modified accordingly. 
In the following, we also make use of partial integrals $I_E^{(K)} (R)$ 
obtained from Eq.~(\ref{6}) with the different terms in Eq.~(\ref{7}). 
The sum of the $I_E^{(K)} (R)$ is $I_E (R)$. 
\section{Potentials}
\label{sec:pot}
The deuteron wave function $u_d$ is calculated with the central Minnesota interaction \cite{TLT77} 
(see Ref.~\cite{DDB03} for details). 
An energy $E_d = -2.202$ MeV is obtained. 

The $^9\textrm{Li}+n+n$ wave function is calculated with the $^9\textrm{Li}+n$ potential P2 of Ref.~\cite{TZ-94} 
and the $nn$ Minnesota interaction with exchange parameter $u=1$.  
In order to fit the binding energy of $^{11}$Li, 
the P2 interaction is multiplied by a parameter \cite{Ba-97}. 
The values 0.992 and 0.9965 provide $S_{2n} = 0.307$ and 0.376 MeV, respectively. 
The $s$-wave scattering length is then slightly modified from $-25.4$ fm 
to $-19.0$ or $-22.2$ fm, respectively. 
Potential P2 contains a forbidden state in the $s$ wave which is eliminated 
with the pseudopotential method \cite{KP-78}. 
Forbidden states need not be eliminated in two-body systems as they do not affect 
scattering properties. 
Their presence leads to more realistic wave functions for the relative motion. 
In three-body systems however, forbidden states must be eliminated because otherwise 
they would unrealistically contribute to the binding energy. 
The pseudopotential moves them to a high energy without affecting the other properties 
of the two-body potentials. 

A $^9\textrm{Li}+d$ optical potential has been obtained by fitting elastic scattering data 
at a c.m.\ energy of 3.86 MeV \cite{JAV05}. 
The real part of this potential does not display any resonance below the Coulomb barrier. 
Such a resonance has been observed in several channels at the excitation energy 
$18.15 \pm 0.15$ MeV \cite{BGG97}, i.e.\ at the c.m.\ energy $0.25 \pm 0.15$ MeV 
above the $^9\textrm{Li}+d$ threshold. 
As shown below, this resonance is crucial to explain the order of magnitude of the 
$\beta$ delayed deuteron decay of $^{11}$Li. 
The potential of Ref.~\cite{JAV05} is thus not useful here. 
Its real part provides three bound states. 
When the depth of its real part is reduced from 104.6 MeV to 89 MeV, 
the upper bound state becomes a resonance near the experimental value. 
However the agreement with the elastic scattering experiment is then lost. 
 
We approximate the $^9\textrm{Li}+d$ potential by expressions based on simple physical arguments 
derived from a microscopic cluster model interpretation. 
(i) At short distances, $^9$Li and deuteron can form a bound state in the $s$ wave. 
This bound state has the same parity as the $^9$Li core, i.e.\ a negative parity. 
We thus impose to the potential to reproduce the energy of the $1/2^-$ excited state of $^{11}$Be 
at an excitation energy of 0.320 MeV. 
This means that our potentials will have a bound state near $-17.6$ MeV. 
(ii) In the microscopic cluster model, the $^9\textrm{Li}+d$ system possesses a forbidden state in the $s$ wave. 
The role of such a state can be simulated by a potential deep enough to contain an unphysical bound state 
below the physical bound state in order to simulate the correct node structure of the scattering wave function. 
(iii) The $^{11}$Be nucleus displays a resonance around 0.25 MeV above the $^9\textrm{Li}+d$ threshold.  
\begin{figure}[htb]
\begin{center}
\includegraphics[width=7cm]{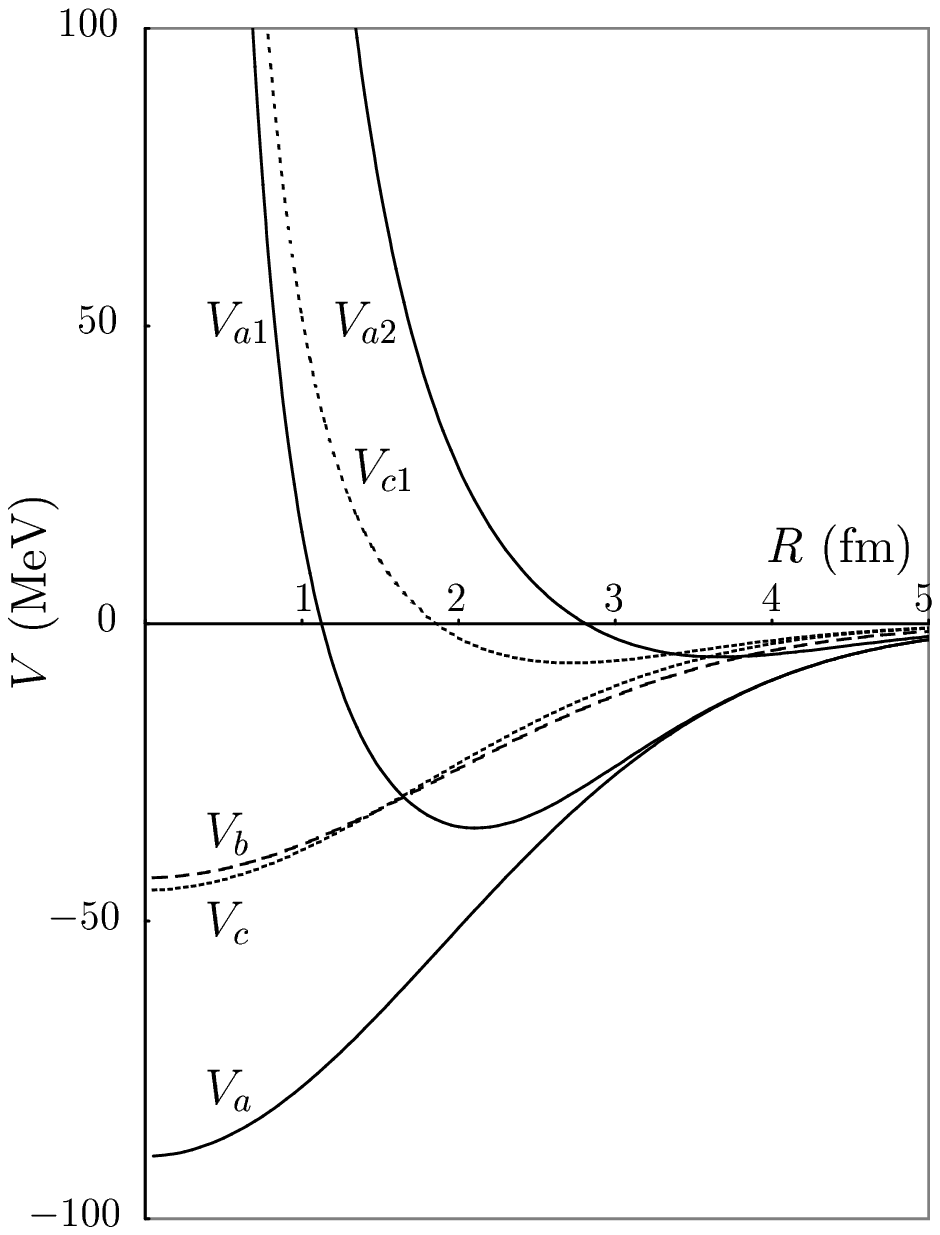}
\end{center}
\caption{Potential $V_a$ and phase-equivalent potentials $V_{a1}$ and $V_{a2}$ (full lines); 
potential $V_b$ (dashed line); 
potential $V_c$ and phase-equivalent potentials $V_{c1}$ (dotted lines). 
\label{Fig1}}
\end{figure}

As in Ref.~\cite{OS-95}, we consider simple Gaussian potentials parametrized as 
\beq
V (R) = - V_0 \exp(-\alpha R^2).
\eeqn{8}
The choice of a Gaussian form factors restricts the number of parameters. 
A Coulomb term $3e^2\, \textrm{erf}(\beta r)/r$ with $\beta =0.71$ fm$^{-1}$ 
(scaled from 0.75 fm$^{-1}$ in the $\alpha+d$ case) is added to all potentials. 
The Coulomb barrier is located between 0.55 and 0.6 MeV. 

\begin{figure}[thb]
\begin{center}
\includegraphics[width=7cm]{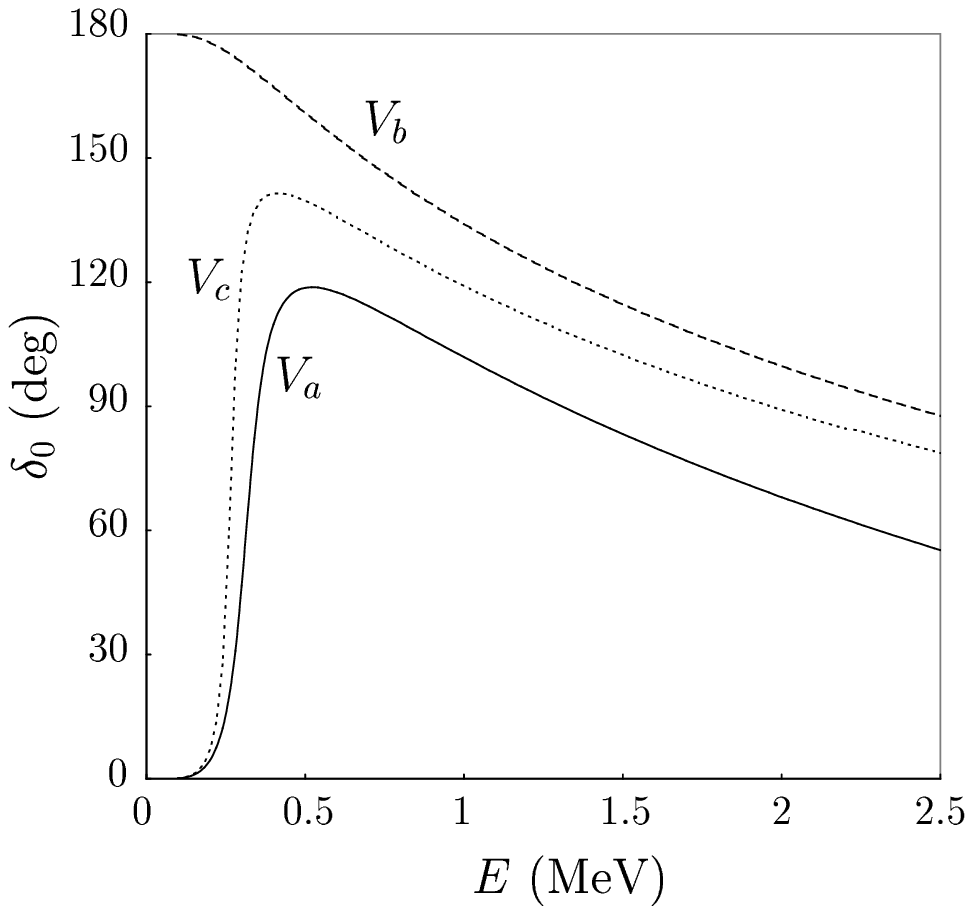}
\end{center}
\caption{$s$-wave $^9\textrm{Li}+d$ phase shifts obtained with potentials $V_a$ (and $V_{a1}$, $V_{a2}$, 
full line), $V_b$ (dashed line) , and $V_c$ (and $V_{c1}$, dotted line). 
\label{Fig2}}
\end{figure}

Potential $V_a$ with $\alpha = 0.14$ fm$^{-2}$ and $V_0 = 89.5$ MeV (see Fig.~\ref{Fig1}) 
has a bound state at energy $-17.63$ MeV and a forbidden state at $-52.42$ MeV. 
The width of the Gaussian form factor has been chosen in such a way 
that the potential also verifies criterion (iii). 
The corresponding phase shift is displayed as a full line in Fig.~\ref{Fig2}. 
A resonance appears at about 0.33 MeV with a width of about 0.1 MeV. 
The spin and parity of this resonance should be $3/2^-$. 
Its width is smaller than the experimental width derived in Ref.~\cite{BGG97} which however 
largely exceeds the Wigner limit and is therefore questionable. 

In the $^6$He case, a forbidden state plays a crucial role in the reproduction of the 
experimental order of magnitude. 
In order to study the role of the forbidden state here, 
we perform pairs of supersymmetric transformations \cite{Ba-87} 
in order to remove it from $V_a$ while keeping the other bound state 
and the $s$-wave resonance and phase shift. 
The resulting phase-equivalent potential denoted as $V_{a1}$ 
exhibits a strong repulsive core (see Fig.~\ref{Fig1}). 
The physical bound state of $V_{a1}$ is then removed by another pair of transformations 
leading to the phase-equivalent potential $V_{a2}$ without any bound state. 
Both potentials $V_{a1}$ and $V_{a2}$ provide the same $s$-wave phase shift as $V_a$ in Fig.~\ref{Fig2}. 

We also consider other Gaussian potentials. 
Potential $V_b$ with the same range as $V_a$ but $V_0 = 42.7$ MeV has its ground state at $-17.63$ MeV. 
This potential possesses a weakly bound state near $-0.184$ MeV in place of a resonance. 
Potential $V_c$ is quite similar to $V_b$ but differs from it by the fact that it possesses a resonance at 0.28 MeV 
in addition to a bound state at $-17.66$ MeV, with $\alpha = 0.161$ MeV and $V_0 = 44.8$ MeV. 
Removing the bound state leads to the phase-equivalent potential $V_{c1}$ with a repulsive core. 
These potentials are compared in Fig.~\ref{Fig1} 
and their phase shifts are displayed in Fig.~\ref{Fig2}. 
One observes that the phase shift of potential $V_c$ has the same shape as the phase shift 
of $V_a$ and also displays the expected resonance but at a lower energy. 
On the contrary, the phase shift of potential $V_b$ is monotonic. 

\begin{figure}[thb]
\begin{center}
\includegraphics[width=10cm]{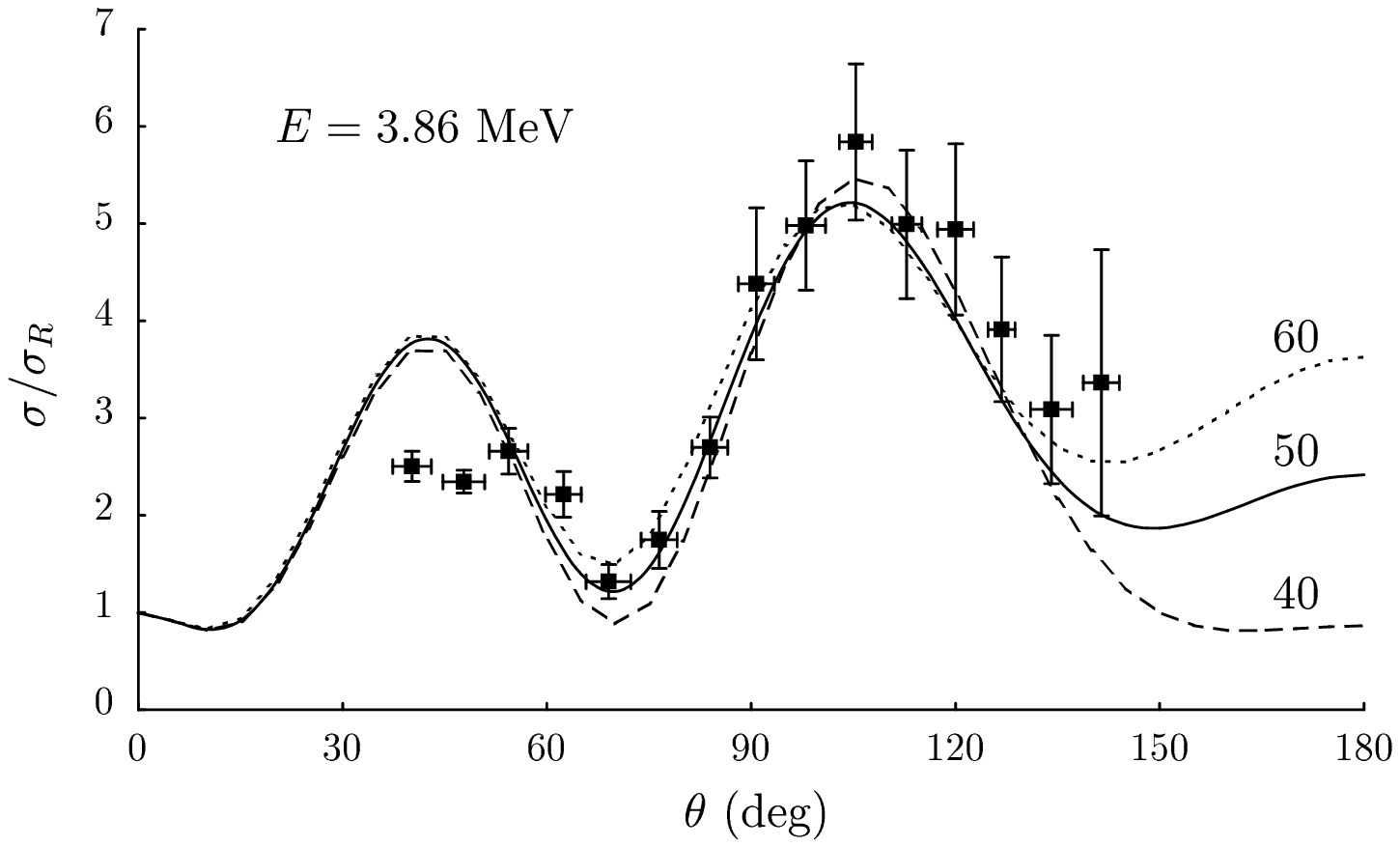}
\end{center}
\caption{Ratio of the elastic cross section to the Rutherford cross section 
at a c.m.\ energy of 3.86 MeV for potential (\ref{9}) with $\alpha = 0.14$ fm$^{-2}$, $V_0 = 89.5$ MeV, 
and $W_0 =40$ (dashed line), 50 (full line), and 60 (dotted line) MeV. 
Experimental data are from Ref.~\cite{JAV05}. 
\label{Fig2a}}
\end{figure}

These potentials are compatible with the elastic scattering data of Ref.~\cite{JAV05} 
if some surface absorption is added. 
Without absorption, even the order of magnitude of the cross section is incorrect 
beyond 70 degrees. 
We use the simple optical potential 
\beq
V_{\rm opt} (R) = -(V_0 + i W_0 \sqrt{\alpha} R) e^{-\alpha R^2},
\eeqn{9}
where the imaginary part is proportional to the derivative of the real part. 
The ratio of the elastic cross section to the Rutherford cross section 
is compared with experiment in Fig.~\ref{Fig2a} for an optical potential 
based on the parameters of $V_a$. 
The only free parameter is thus $W_0$. 
One observes that a fair agreement with the data is obtained for $W_0 = 50 \pm 10$ MeV, 
except around the first maximum near 40 degrees. 
The present real part is thus compatible with elastic scattering. 
\section{Gamow-Teller integrals}
\label{sec:GTI}
\begin{figure}[htb]
\begin{center}
\includegraphics[width=7cm]{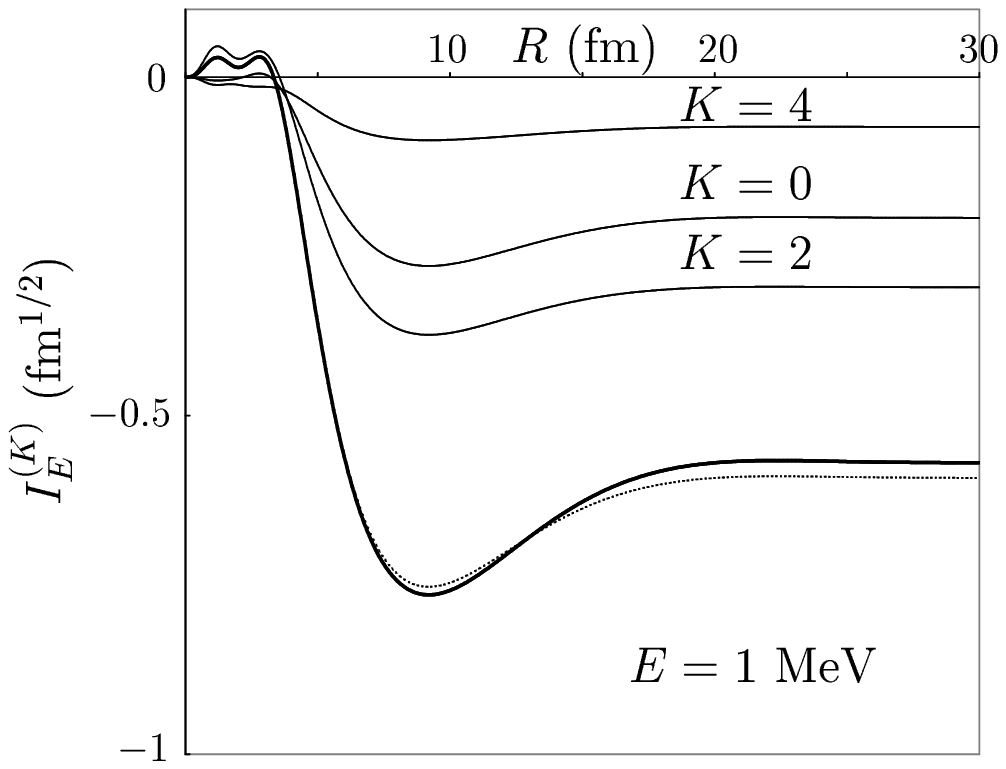}
\end{center}
\caption{Partial integrals $I_E^{(K)}(R)$ for $K = 0$, 2, and 4 
at the energy $E=1$ MeV for potential $V_a$. 
The sum of the three components $K = 0$, 2, and 4, (dotted line) 
and the converged sum $I_E (R) = \sum_K I_E^{(K)}(R)$ [Eq.~(\ref{6})] 
(lowest full line) are also displayed. 
\label{Fig3}}
\end{figure}

The integrals $I_E^{(K)}(R)$ calculated with potential $V_a$ are displayed in Fig.~\ref{Fig3} 
as a function of $R$ for different $K$ values at energy 1 MeV. 
The convergence of $\sum_K I_E^{(K)}(R)$ is reached for $K_{\rm max} = 20$. 
Partial waves $K = 2$, 0, and 4 are strongly dominant, 
although the cumulated contribution of all higher partial waves is not negligible. 
Contrary to the $^6$He case, no important cancellation is encountered here 
because the dominant contributions have the same sign. 
\begin{figure}[ht]
\begin{center}
\includegraphics[width=7cm] {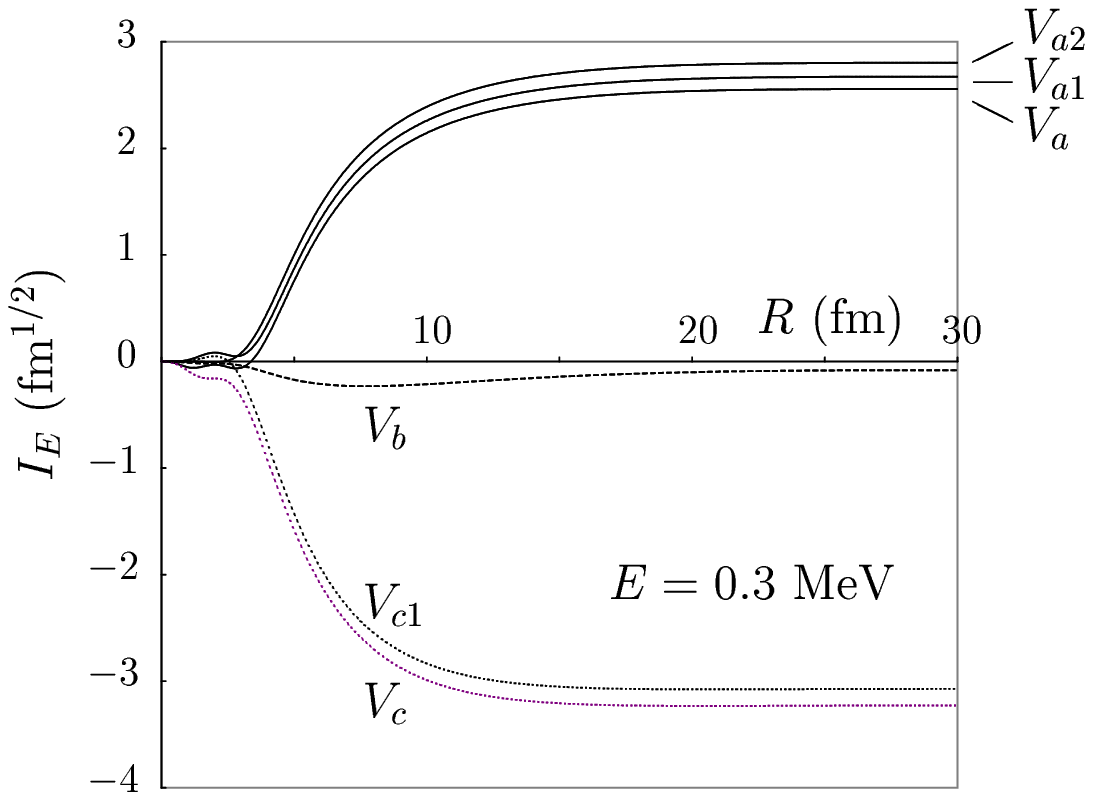}
\end{center}
\begin{center}
\includegraphics[width=7cm] {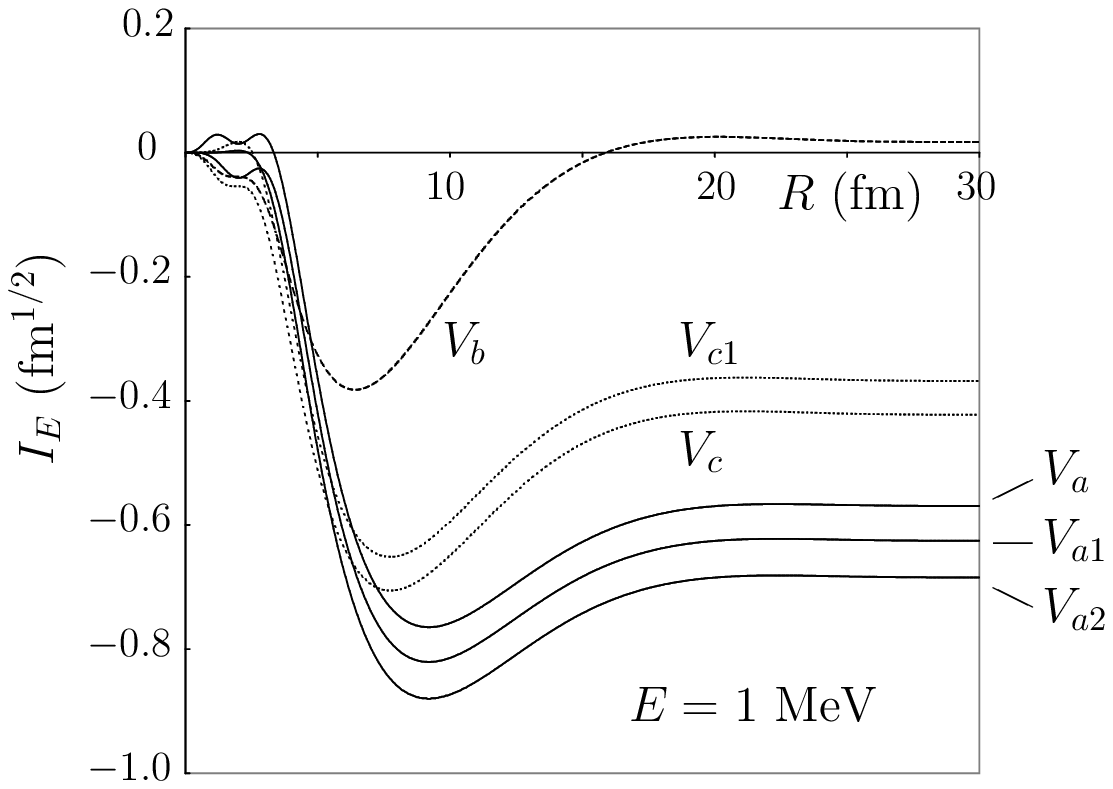}
\end{center}
\caption{Integrals $I_E (R)$ [Eq.~(\ref{6})] at (a) $E=0.3$ MeV 
and (b) 1 MeV, for various potentials. 
\label{Fig4}}
\end{figure}
%

In Fig.~\ref{Fig4}, the integrals $I_E (R)$ are compared 
for the different potentials considered at two energies: 
the near-resonance energy 0.3 MeV and a typical non-resonant energy 1 MeV. 
As shown by Eq.~(\ref{6}), the integral $I_E$ displays a minimum or a maximum 
every time either $u_E$ or $u_{\rm eff}$ vanishes. 
Its behavior depends on the node structure of the scattering 
wave function and thus on the depth of the potential. 
At both energies, all curves present an extremum near 2 fm 
which corresponds to the unique node of $u_{\rm eff}$. 
They present one to three additional nodes corresponding to the 
possible bound states and resonance of the potential. 
However, in spite of their different numbers of bound states and thus of nodes, 
potentials $V_a$, $V_{a1}$, and $V_{a2}$ do not give very different results 
at both energies. 
In all three cases, the amplitude of the integral starts to increase beyond 3 fm 
and reaches a plateau near 20 fm. 
At 1 MeV, it presents a maximum near 9 fm. 
This maximum is at the same location for $V_a$, $V_{a1}$, and $V_{a2}$ 
because phase-equivalent potentials have the same asymptotic behavior 
and thus the same nodes beyond the potential range. 
This situation must be contrasted with the $^6$He case where the cancellation 
enhances tiny differences and where phase-equivalent potentials provide very 
different results \cite{TBD06}. 

The results for potential $V_c$ which satisfies the same physical conditions as $V_{a1}$ 
are very similar because the scattering wave function has the same number of nodes 
and similar locations of these nodes. 
At 0.3 MeV, the integrals have opposite signs for $V_a$ and $V_c$ beyond 4 fm 
because the resonance is below 0.3 MeV for $V_c$ while it is above for $V_a$. 
They have the same sign and similar magnitudes at 1 MeV. 

On the contrary, the results obtained with $V_b$ are very different, even off resonance, 
because the scattering wave function has nodes at quite different locations. 
In particular its node near 8 fm at 0.3 MeV or 6 fm at 1 MeV leads to a cancellation 
similar to that of the $^6$He case. 
We shall see in the next section that this type of result is ruled out by experiment. 
It is important to realize that $V_b$ has the same physical bound state as $V_{a1}$ and $V_c$ near $-17.6$ MeV. 
However, $V_b$ does not reproduce the resonance (see Fig.~\ref{Fig2}). 
\section{Transition probability per time and energy units}
\label{sec:res}
\begin{figure}[hbt]
\begin{center}
\includegraphics[width=10cm]{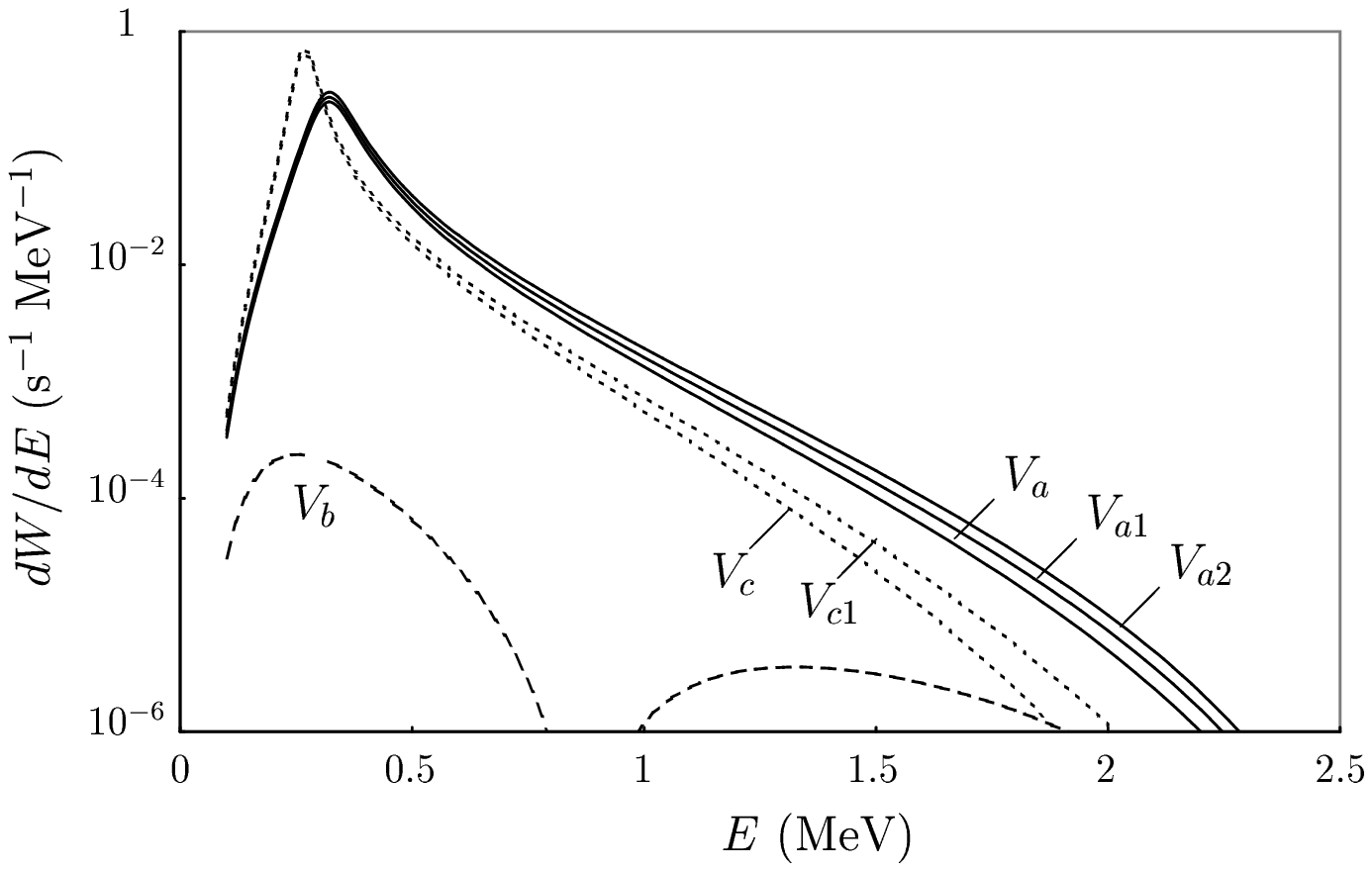}
\end{center}
\caption{Transition probability per time and energy units $dW/dE$ 
of the $^{11}$Li $\beta$ decay into the $^9\textrm{Li}+d$ continuum 
as a function of the relative $^9\textrm{Li}+d$ energy $E$ 
calculated with various $^9\textrm{Li}+d$ potentials. 
\label{Fig5}}
\end{figure}
The transition probability per time and energy units given by Eq.~(\ref{1}) 
is plotted in Fig.~\ref{Fig5} as a function of the relative $^9\textrm{Li}+d$ energy $E$ for the different potentials. 
For the potentials displaying a resonance, the results are qualitatively very similar. 
The shape of the curve does not depend much on the resonance location. 
On the contrary, potential $V_b$ provides results with a much smaller order of magnitude. 
The same situation is observed with other potentials that do not possess a resonance. 

The total transition probabilities per time unit (integrated from 0 or from some cutoff to $Q$) 
corresponding to the various potentials are compared 
with the experimental value in Table \ref{table1}. 
This value is calculated from the experimental branching ratio $\cal R$ by 
\beq
W_{\rm exp} = {\cal R} \ln 2 /t_{1/2} \approx 81.5\, {\cal R}\ \textrm{s}^{-1},
\eeqn{10}
where $t_{1/2}$ is the $^{11}$Li ground-state half life 8.5 ms. 
The experimental branching ratio is $(1.5 \pm 0.2) \times 10^{-4}$ \cite{BGG97}. 
The results at various cutoff values in Table \ref{table1} will be useful 
for comparison with the new experiment \cite{Ra-06}.

All potentials except $V_b$ provide the right order of magnitude but overestimate 
the experimental value of Ref.~\cite{BGG97} by a factor larger than 3. 
Results of successive phase-equivalent potentials differ by about 10 \%.
The main difference between $V_a$ and $V_c$ arises in the cutoff dependence 
which is sensitive to the resonance location. 
\begin{table}[ht]
\caption{Total transition probability per second $W$ (in $10^{-3}$ s$^{-1}$) 
for the $\beta$ decay of $^{11}$Li into $^9\textrm{Li}+d$. 
For each value of the two-neutron separation energy $S_{2n}$ (in MeV), 
the rows correspond to various cutoffs.}
\vspace*{0.3 cm}
\begin{tabular}{clrrrrrrr} 
\hline
$S_{2n}$ & \multicolumn{1}{c}{cutoff} & \multicolumn{1}{c}{$V_a$} 
& \multicolumn{1}{c}{$V_{a1}$} & \multicolumn{1}{c}{$V_{a2}$} & \multicolumn{1}{c}{$V_b$} 
& \multicolumn{1}{c}{$V_c$} & \multicolumn{1}{c}{$V_{c1}$} & \multicolumn{1}{c}{Exp.}                    \\ 
\hline
0.307 & $E>0$     & 38.1 & 42.1   & 46.7   & 0.0718 & 59.7 & 54.1   & $12 \pm 2$ \cite{BGG97} \\
      & $E > 0.3$ & 31.0 & 34.3   & 38.2   & 0.0392 & 22.0 & 19.7   & $        $              \\
      & $E > 0.5$ &  4.7 &  5.4   &  6.2   & 0.0096 &  2.7 &  2.3   & $        $              \\
0.376 & $E>0$     & 31.5 & 34.9   & 39.0   & 0.1014 & 50.8 & 45.8   & $12 \pm 2$ \cite{BGG97} \\
      & $E > 0.3$ & 25.7 & 28.6   & 32.1   & 0.0622 & 19.0 & 17.0   & $        $              \\
      & $E > 0.5$ &  4.0 &  4.6   &  5.3   & 0.0185 &  2.4 &  2.1   & $        $              \\
\hline
\end{tabular}
\label{table1}
\end{table}

Table \ref{table1} also indicates the role of a larger two-neutron separation energy $S_{2n}$. 
This introduces a modification of the $Q$ value and of the $^{11}$Li wave function. 
Except for $V_b$, the transition probabilities are slightly reduced, by about 20 \%. 
This effect is rather weak and does not modify the discussion. 
The $V_b$ variation emphasizes the high sensitivity to weak modifications 
when a cancellation occurs, like in the $^6$He case. 

Let us study the role of the main uncertainties in our theoretical description. 
The first uncertainty concerns the energy location of the resonance. 
The location of the peak in Fig.~\ref{Fig5} affects the total transition probability. 
In Table \ref{table2}, we study the dependence of the transition probability 
on the resonance energy $E_r$. 
To this end, we slightly vary the depth $V_0$ in potential $V_a$. 
This leads to a small violation of our criterion (i), i.e.\ the energy $E_{BS}$ of the physical 
bound state is somewhat modified, but this modification remains acceptable 
in view of our other simplifying assumptions. 
As shown by Table \ref{table2}, $W$ is locally quite sensitive to the resonance energy 
and a slightly higher location would lead to smaller values. 
A higher location of the resonance also reduces the cutoff dependence. 
\begin{table}[ht]
\caption{Dependence of the total transition probability per second $W$ (in $10^{-3}$ s$^{-1}$) 
on the resonance energy $E_r$ (in MeV) calculated with a Gaussian potential with $\alpha = 0.14$ fm$^{-2}$ 
as a function of its depth $V_0$ (in MeV) for various cutoffs. 
The forbidden state energy $E_{FS}$ and the physical bound state energy $E_{BS}$ are also displayed.}
\vspace*{0.3 cm}
\begin{tabular}{ccccrrr} 
\hline
 $V_0$ & $E_{FS}$ & $E_{BS}$ & $E_r$  & \multicolumn{3}{c}{$W$}  \\
       &          &          &        & $E>0$ &$E>0.3$&$E>0.5$\\
\hline
90.8   & $-53.44$ & $-18.26$ &  0.25  & 47.5  & 10.7  &  2.1  \\
90.1   & $-52.89$ & $-17.92$ &  0.30  & 42.1  & 24.6  &  3.3  \\
89.5   & $-52.42$ & $-17.63$ &  0.35  & 38.1  & 31.0  &  4.7  \\
89.0   & $-52.02$ & $-17.38$ &  0.40  & 35.2  & 31.0  &  6.2  \\
88.5   & $-51.63$ & $-17.15$ &  0.46  & 32.5  & 29.6  &  8.0  \\
\hline
\end{tabular}
\label{table2}
\end{table}

\begin{figure}[htb]
\begin{center}
\includegraphics[width=10cm]{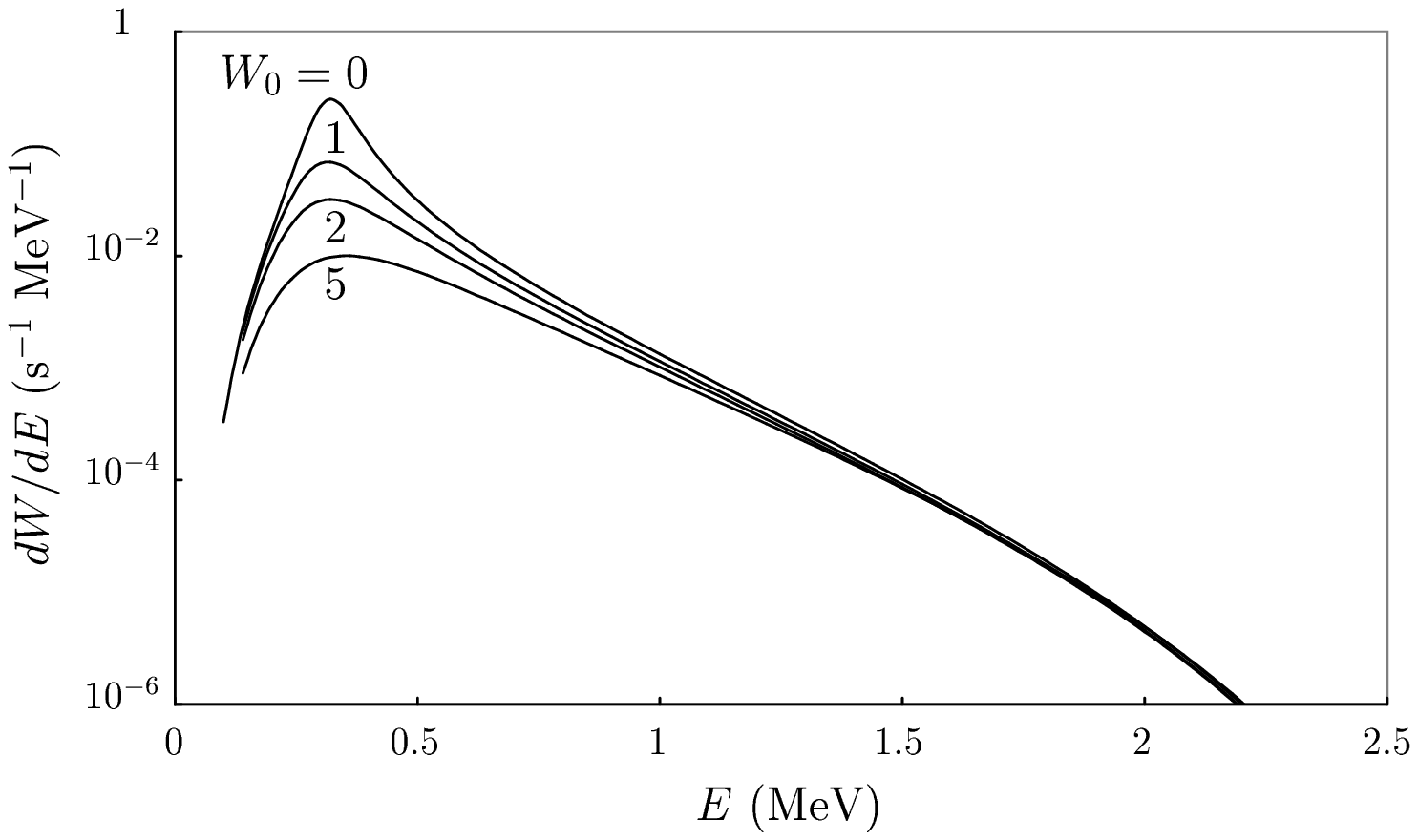}
\end{center}
\caption{Transition probability per time and energy units $dW/dE$ 
of the $^{11}$Li $\beta$ decay into the $^9\textrm{Li}+d$ continuum 
as a function of the relative $^9\textrm{Li}+d$ energy $E$ 
calculated with various values of the surface absorption strength $W_0$ (in MeV) in Eq.~(\ref{9}). 
\label{Fig6}}
\end{figure}

Another effect, not encountered in the $^6$He case, arises from the fact 
that several channels are open below the $^9\textrm{Li}+d$ channel, 
the lowest one being the $^{10}\textrm{Be}+n$ channel. 
Transfer towards these channels is possible at all energies 
but should be rather weak below the Coulomb barrier. 
The magnitude of the surface absorption into these channels has been derived from experimental scattering data 
in Fig.~\ref{Fig2a} but at an energy 3.86 MeV much higher than the Coulomb barrier. 
Therefore we restrict the discussion to small values of $W_0$ in Eq.~(\ref{9}) 
(this parameter should probably depend on energy but we neglect this effect here).  
One observes in Fig.~\ref{Fig6} that the role of the resonance is strongly reduced 
even by a weak absorption. 
On the contrary, the results above 1 MeV are not much affected. 
The energy dependence of $dW/dE$ becomes weaker when absorption increases. 

The total transition probabilities per second $W$ calculated with potential (\ref{9}) in Eq.~(\ref{5}) 
are displayed in Table~\ref{table3} with $\alpha$ and $V_0$ as in potential $V_a$ 
for several fixed values of the surface absorption constant $W_0$. 
One observes that a much smaller absorption than in Fig.~\ref{Fig2a} leads to a strong 
reduction of the transition probability. 
As explained by Fig.~\ref{Fig6}, absorption leads to a weaker dependence on the cutoff. 
\begin{table}[ht]
\caption{Dependence of the total transition probability per second $W$ (in $10^{-3}$ s$^{-1}$) 
on the surface absorption strength $W_0$ (in MeV), calculated with potential $V_a$ as real part
for various cutoffs.}
\vspace*{0.3 cm}
\begin{tabular}{crrr} 
\hline
 $W_0$ &  $E>0$  & $E>0.3$ & $E>0.5$ \\
\hline
0      & 38.1  & 31.0  &  4.7  \\
1      & 16.7  & 13.0  &  3.5  \\
2      &  9.9  &  7.8  &  2.7  \\
5      &  4.4  &  3.6  &  1.8  \\
\hline
\end{tabular}
\label{table3}
\end{table}
\section{Conclusions}
\label{sec:conc}
In the present work, we studied the $\beta$-decay process of the $^{11}$Li halo nucleus 
into the $^9\textrm{Li}+d$ continuum in the framework of a three-body model. 
Three-body hyperspherical bound-state wave functions on a Lagrange mesh and two-body $^9\textrm{Li}+d$ 
scattering wave functions have been used. 
For the calculation of the $\beta$-decay transition probabilities per time and energy units, 
several $^9\textrm{Li}+d$ potentials were employed. 

Some $^9\textrm{Li}+d$ potentials are physically inspired by a microscopic cluster picture 
and involve a forbidden state and a physical bound state simulating the $1/2^-$ excited state of $^{11}$Be. 
A resonance occurs in the $s$ wave at about the experimental energy. 
For potentials of this family, the transition probability per time unit is weakly sensitive to the potential choice. 
However a potential without this resonance fails to reproduce even the order of magnitude 
of the transition probability. 
The high sensitivity of the delayed $\beta$ decay of $^6$He due to a cancellation in the Gamow-Teller 
matrix element does not occur here. 
This is emphasized by using phase-equivalent potentials differing by their number of bound states: 
they give very different results for $^6$He and very similar results for $^{11}$Li. 

The theoretical result is sensitive to the $^{11}$Li separation energy (about 20 \% 
if $S_{2n}$ is increased by about 70 keV). 
It is also sensitive to the location of the resonance. 
A more accurate experimental determination of the location of this resonance (with a precise definition 
of the resonance energy for such a broad resonance) would be very useful. 
Elastic scattering data on $d(^9\textrm{Li},^9\textrm{Li})d$ at an energy close to the Coulomb barrier 
extending up to backward angles 
might also help reducing the uncertainty on the parameters of the optical potential for this collision. 

The calculated transition probability overestimates the experimental result of Ref.~\cite{BGG97} 
by a factor larger than 3. 
The overestimation can be reduced by modifying the resonance location and/or by introducing absorption 
removing flux from the $^9\textrm{Li}+d$ final channel. 
\section*{Acknowledgments}
We thank H.~Jeppesen, A.M. Moro, and R.~Raabe for information about their experiments. 
This text presents research results of the Belgian program P5/07 on 
interuniversity attraction poles initiated by the Belgian-state 
Federal Services for Scientific, Technical and Cultural Affairs (FSTC).
P.D. and E.M.T. acknowledge the support of the National Fund for Scientific Research (FNRS), Belgium. 
\end{document}